\documentclass[12pt]{article}

\newread\testifexists
\def\GetIfExists #1 {\immediate\openin\testifexists=#1
    \ifeof\testifexists\immediate\closein\testifexists\else
    \immediate\closein\testifexists\input #1\fi}

\usepackage{gthstyle}\usepackage{amsfonts}
\usepackage{amssymb}
\immediate\openin\testifexists=e
    \ifeof\testifexists\immediate\closein\testifexists\else
    \immediate\closein\testifexists\input e\fipsf

\def\bbs#1{\setbox0=\hbox{$\tt #1$}  \copy0\kern-\wd0\kern .15em\copy0}

\def\bbf#1{\setbox0=\hbox{$#1$} \kern-.025em\copy0\kern-\wd0
        \kern.05em\copy0\kern-\wd0 \kern-.025em\raise.0433em\box0}

      \def\b{\beta}         
\def\d{\delta}        
          \def\l{\lambda}     \def\L{\Lambda}
\def\m{\mu}         \def\f{\phi}    \def\F{\Phi}        
\def\n{\nu}         \def\j{\psi}    
\def\r{\varrho}     \def\s{\sigma}

\def\HH{{\cal H}} \def\LL{{\cal L}} \def\OO{{\cal O}}\def\NN{{\cal
N}}
\def\pa{\partial} \def\ra{\rightarrow}

 \def\bel{$\circ$}
\def\dd{{\rm d}}  \def\bra{\langle}   \def\ket{\rangle}

\def\fract#1#2{{\textstyle{#1\over#2}}}
\def\ffract#1#2{\raise .3 em\hbox{$\scriptstyle#1$}\kern-.25em/
                \kern-.2em\lower .2 em \hbox{$\scriptstyle#2$}}

\def\half{\fract12} \def\quart{\fract14} 

\def\part#1#2{{\partial#1\over\partial#2}}

\def\iss{\ =\ }

\newcommand{\tl}[1]{\tilde{#1}}

\newcommand{\Tr}{{\mbox{Tr}}\,}
\newcommand{\be}{\begin{eqnarray}}
\newcommand{\ee}{\end{eqnarray}}
\newcommand{\eqn}[1]{(\ref{#1})}
\newcommand{\nn}{\nonumber\\}

\newcommand{\bi}[1]{\begin{itemize}\item[#1]}
\newcommand{\itm}[1]{\item[#1]}
\newcommand{\ei}{\end{itemize}}

\newcommand{\fn}{\footnote}

\newcommand{\newsec}[1]{\section{#1}\setcounter{equation}{0}}

\newcommand{\eel}[1]{\label{#1}\end{eqnarray}}\newcommand{\crl}[1]{\label{#1}\\ }

\begin{document}
\begin{titlepage}

\title{\normalsize \hfil  ITP-UU-05/17 \hfil SPIN-05/13
 \hfil {\tt gr-qc/0504120}\\ \vskip 10mm \Large\bf
THE HOLOGRAPHIC MAPPING OF THE STANDARD MODEL ONTO THE BLACK HOLE
HORIZON\\ \emph{Part I: Abelian Vector Field, Scalar Field and BEH
Mechanism}}

\author{Gerard 't~Hooft}
\date
 {\normalsize Institute for Theoretical Physics,
Utrecht University \\ and
\medskip \\ Spinoza Institute, Postbox 80.195, 3508 TD
Utrecht, the Netherlands \smallskip \\ e-mail: \tt
g.thooft@phys.uu.nl \\ internet: \tt
http://www.phys.uu.nl/\~{}thooft/}

\maketitle

\begin{quotation} \noindent {\large\bf Astract } \medskip \\
 Interactions between outgoing Hawking particles and ingoing matter are determined by gravitational forces and Standard Model
interactions. In particular the gravitational interactions are responsible for the unitarity of the scattering against the
horizon, as dictated by the holographic principle, but the Standard Model interactions also contribute, and understanding
their effects is an important first step towards a complete understanding of the horizon's dynamics. The relation between
in- and outgoing states is described in terms of an operator algebra. In this paper, the first of a series, we describe the
algebra induced on the horizon by \(U(1)\) vector fields and scalar fields, including the case of an Englert-Brout-Higgs
mechanism, and a more careful consideration of the transverse vector field components.
\end{quotation}


\end{titlepage}

\eject
\def\inn{\mathrm{in}}\def\out{\mathrm{out}}\def\intt{\mathrm{int}}
\newsec{Introduction: The black hole scattering matrix}

A growing consensus seems to have been reached concerning the hypothesis that the contribution of black holes to particle
scattering phenomena must be described by a unitary scattering operator\cite{GtH85}\cite{tHSW}\cite{GtH96}\cite{SWH04}, but
the agreement is not universal, and indeed, it is not at all obvious\cite{WUpc}. The author's personal arguments favoring
unitarity contain elements that are not at all agreed upon by any of his colleagues; since these are not essential, we defer
them to Appendix {\ref{unitarity.app}} of this paper. Whatever one's views are, unitarity, assumed to hold regardless the
presence of black holes, is an essential starting point of this paper, and its sequels.

The unitarity assumption is applied to the region within the Planckian
domain surrounding the horizon. The Hawking particles leaving that domain,
will subsequently be affected by the ingoing particles by interactions
that we assume to be known. There are two types of interactions that we
can take into account: \bi{\bel} those caused by \emph{perturbative}
quantum gravity, and \itm{\bel} those described by some renormalizable
quantum field theory, to be referred to as the ``Standard Model". \ei The
gravitational interactions are taken to be perturbative because we are
looking at scales where the higher order corrections may already be
assumed to be negligible. They, as well as the Standard Model
interactions, can be regarded as ``final state interactions", taking place
well outside the Planckian regime. To handle these final state
interactions, the so-called \emph{eikonal approximation}\cite{AbItz} seems
to be appropriate. Outgoing particles scatter against ingoing ones at high
center of mass energy and low values for the exchanged momentum. Possibly,
this technique can be used in an entirely general sense, to obtain all
effects due to Standard Model interactions, but, in this paper, we choose
a different approach.

The final state interactions are not as innocuous as they might seem. The
gravitational ones, in particular, have a divergent Rindler time
dependence, so that their effects are by no means small. Indeed, they
modify the local spectrum so much that the unitarity assumption can only
be regarded in harmony with the presence of these effects. Together, the
interactions produce a new boundary condition at the horizon, in the form
of an operator algebra, which we wish to study. The picture one obtains
appears to be self-consistent, apart from the fact that the transverse
components of the gravitational interactions are difficult to handle.
Ignoring the transverse components is tantamount to neglecting the finite
size effects in the transverse directions on the horizon, so that it
should not come as a surprise that one then does not recover the desired
area law for the statistical entropy.

The transverse components of the gravitational field resemble somewhat the "Standard Model" interactions, and this is why we
are taking a closer look at those first.

The result of these investigations is that operators describing in- and outgoing particles in the vicinity of the horizon
are described by an operator algebra. The algebra represents all interactions between in- and outgoing objects up to some
distance \(1/\m\) from the horizon (according to the local metric), where \(\m\) is the scale parameter at which the
Standard Model characteristics have been taken into account.

We consider a small section of the horizon of a black hole, while the hole itself is taken to be large, so that this region
of space-time is adequately described by the Rindler metric. Of course we are in three space dimensions and one time. In
locally flat coordinates, the past horizon is described by a light cone coordinate \(x^+=U_\inn(\tl\s)\) and the future
horizon is \(x^-=U_\out(\tl\s)\), where \(\tl\s=(\s_x,\,\s_y)\) are the transverse coordinates of the horizon. Since
(transverse) gravitational interactions are ignored from here on, there will be no back reaction on the metric of the
horizon. Therefore, the horizon's metric can be taken to be a fixed background. In this paper, we limit ourselves to the
transverse components of the metric being \(g_{ij}=\d_{ij}\), so that the \(\tl\s\) are just cartesian coordinates.
Generalization to any other choice for the background metric will be straight-forward.

The main result described in Ref.\cite{GtH96}, is that, in the approximation where transverse gravity is neglected, the
ingoing particles can be described by a momentum distribution \(P_\inn(\tl\s)\) and the outgoing ones by \(P_\out(\tl\s)\),
and the following algebra is obtained: \be \big[\,
P_\inn(\tl\s),\,U_\inn(\tl\s')\,\big]&=&\big[\,P_\out(\tl\s),\,U_\out(\tl\s')\,\big]\iss -i\d^2(\tl\s-\tl\s')\crl{pucomm}
\big[\,P_\inn(\tl\s),\,P_\inn(\tl\s')\,\big]&=&\big[\,U_\inn(\tl\s),\,U_\inn(\tl\s')\,\big]\iss
0\ ,\\
\big[\,P_\out(\tl\s),\,P_\out(\tl\s')\,\big]&=&\big[\,U_\out(\tl\s),\,U_\out(\tl\s')\,\big]\iss 0\ ,\\
P_\out(\tl\s)\iss\tl\pa_\s^2\, U_\inn(\tl\s)\!\!&,&\ P_\inn(\tl\s)\iss -\tl\pa_\s^2\, U_\out(\tl\s)\ .\eel{gravalgebra} This
is a gravitational effect; Newton's constant is normalized to \be 8\pi G=1\ .\eel{newtonnorm} \emph{If} we assume that all
states are unambiguously identified by specifying the momentum distribution \(P(\tl\s)\), we find that we can either choose
a basis where \(P_\inn(\tl\s)\) is specified, or a basis where \(P_\out(\tl\s)\) is specified. A unitary transformation
connects these two basis sets, and so we have a unitary \(S\) matrix.

Since transverse gravity was not taken into account, we could not have expected that the spectrum exhibits only a finite
number of states for every given segment of the horizon. The area law is not yet there, because the transverse metric of the
horizon is not subjected to gravitational interactions. Before handling transverse gravity, we should obtain experience in
deducing such algebras from interacting field theories; this is why, in this paper, we now concentrate on the Standard Model
interactions. One of these was already handled in Ref.\cite{GtH96}: the Maxwell field interaction. It provides for
additional operators, obeying their own algebra. Continuing this way, we get a more precise operator algebra, which should
pave the way to complete control of the horizon.

The phrase ``Holography" in the title refers to the fact that the variables on the horizon appear to span a Hilbert space
associated to a two-dimensional surface, whereas all in-going and out-going particles in the vicinity of the black hole
should be included in these degrees of freedom\cite{GtH93}. This, we actually relax somewhat. At a given renormalization
scale \(\m\) of the Standard Model, we ignore infrared divergences at much lower energy scales. The infra-red cut-off then
amounts to ignoring degrees of freedom far away from the horizon. This way, one may regard our horizon algebra as a
\emph{boundary condition} on the horizon.

\newsec{\(U(1)\)-vector fields}\label{vector.sec} The first Standard Model effect to be considered is that of a \(U(1)\)
gauge field, \(A_\m(x)\), see also Ref.\cite{GtH96}. It suits us well to illustrate our general procedure.  In 3+1
dimensions, the Lagrangian is
 \be S=\int\dd^4x \LL(x)\ ;\qquad \LL(x)=-\quart F_{\m\n}F_{\m\n}
 \ . \eel{MaxwL}
Given the charge distribution \(\r_\inn(\tl\s)\) of the ingoing particles, we find them to generate a vector field
\(A_\m(x)\), which becomes singular on the past horizon. By limiting ourselves to this singular contribution, we select out
the effects of the horizon itself that substitute as a boundary condition relating in- and outgoing objects. The particles
that went in long ago, affect the particles that emerge in the late future. The \(U(1)\) field generated by them takes the
form of a distribution along the past horizon, as in the case of the generation of \v{C}erenkov radiation. Before and behind
the past horizon, the vector field is a pure gauge, but on the horizon these pure gauge parameters \(\L(\tl\s)\) are making
a finite jump. This gauge jump is found to be
 \be\L(\tl\s)=-\int\dd^2\tl\s'\,f(\tl\s-\tl\s')\r_\inn(\tl\s')\ ,
 \qquad\tl\pa_\s^2\,f(\tl\s)=-\d^2(\tl\s)\ . \eel{gaugejump}
On the past horizon itself, the vector field is not a pure gauge but stays finite. Therefore, the gauge jump \eqn{gaugejump}
has a physically relevant effect on any charged outgoing particle: it performs a gauge rotation of the form \be
e^{iQ_\out\L(\tl\s)}\ . \eel{gaugetrfout} In total, the combined wave function of all outgoing particles undergoes a gauge
rotation \be e^{-i\int\dd^2\tl\s\int\dd^2\tl\s'\,f(\tl\s-\tl\s')\r_\inn(\tl\s)\r_\out(\tl\s)} \ ,\eel{gaugetrtot} with which
we should multiply the amplitudes generated by the algebra \eqn{pucomm}-\eqn{gravalgebra}.

Introducing operators \(\l(\tl\s)\) defined by \be \r_{\inn\atop\out}(\tl\s)\dd^2\tl\s&=&-i\pa/\pa\l_{\inn\atop\out}
(\tl\s)\ , \crl{rdl} \big[\r_\inn(\tl\s),\l_\inn(\tl\s')\big]&=& \big[\r_\out(\tl\s),\l_\out(\tl\s')\big]\iss
-i\d^2(\tl\s-\tl\s')\ , \eel{rlcomm} we find that \be\bra\{\l(\tl\s)\}|\{\r(\tl\s)\}\ket=\NN e^{i\int\dd^2\tl\s\,
\r(\tl\s)\l(\tl\s)}\ , \eel{rlampl} both for the in-states and the out-states, where \(\NN\) is a normalization factor.
Comparing this with Eq.\eqn{gaugetrtot}, one gets \be \l_\inn(\tl\s)&=&-\int\dd^2\tl\s'\,f(\tl\s-\tl\s')\r_\out(\tl\s')\ ,
\crl{rlinout} \l_\out(\tl\s)&=&\int\dd^2\tl\s'\,f(\tl\s-\tl\s')\r_\inn(\tl\s') \ ,\eel{rloutin} or \be
\r_\inn(\tl\s)=-\tl\pa_\s^2\,\l_\out\ , \qquad \r_\out(\tl\s)=\tl\pa_\s^2\,\l_\inn\ ;\crl{rinlout}
\big[\,\r_\inn(\tl\s),\,\r_\out(\tl\s')\,\big]=-i\tl\pa_\s^2\d(\tl\s-\tl\s') \ . \eel{inoutcomm} Note, that electric charge
is quantized in multiples of the electric charge unit \(e\). therefore, the phase field \(\l(\tl\s)\) is only well-defined
\emph{modulo} \(2\pi/e\). Thus, the physically relevant field in Eq.~\eqn{rlampl} is \(e^{ie\l(\tl\s)}\,\), rather than
\(\l(\tl\s)\).

This is the main part of, but not yet the complete algebra generated on a flat horizon by a \(U(1)\) gauge field whose
Lagrangian in Minkowski space-time is \eqn{MaxwL}. Note that we have not yet imposed local gauge invariance on the horizon.
All we have is a global symmetry, \(\L(\tl\s)\ra\L(\tl\s)+\L_0\), where the constant field \(\L_0\) is the generator. We
return to this subject in Section \ref{transverse.sec}.

For future use, it is important to find the relation between the algebra \eqn{rlinout}---\eqn{rloutin} and the Lagrangian
\eqn{MaxwL}. We write\fn{Sign conventions in our publications are not uniform. Our present summation convention is \(X_\m
Y_\m=X_+Y_-+X_-Y_++\tl X\tl Y\).}  \be \LL(x)=-\quart F_{\m\n}F_{\m\n}-J_\m(x)A_\m(x)\ , \eel{MaxwLJ} and, concentrating
first on the contribution of the ingoing particles, \be J_+(x)=\r_\inn(\tl\s)\d(x^+)\ ,\qquad J_-(x)=\tl J(x)=0\ .
\eel{insource} In the Lorentz gauge, the vector field obeys \be\pa_+ A_-+\pa_-A_++{\tl\pa}\tl A&=&0\ \nn
2\pa_+\pa_-A_\m+{\tl\pa}^2 A_\m&=& J_\m\ , \eel{Maxwequ} so that \be A_+(x)&=&\d(x^+)\L(\tl\s)\ , \qquad A_-=\tl A=0\ ,\nn
\L(\tl\s)&=&-\int\dd^2\tl\s'\,f(\tl\s-\tl\s')\r_\inn(\tl\s')\iss\l_\out(\tl\s)\ . \eel{AMaxw}

The outgoing particle wave functions are then gauge rotated accordingly. The same, \textit{mutatis mutandis}, happens for
the ingoing particles.

\newsec{Scalar fields}\label{scalar.sec}

Consider now a set of scalar fields \(\F^i(x)\) in the Standard model. The Lagrangian may be taken to be
\be\LL(x)=-\half(\pa\F^i)^2-V(\mathbf\F)\ . \eel{scalarL} Their effect on the algebra is nearly trivial, which is why we
discussed the vector case first. What has to be observed is, that a scalar field that lives on the plane \(x^+=x^-=0\)
defining the horizon, remains constant under a Rindler time boost, \be x^+\ra \l\, x^+ ,\qquad x^-\ra x^-/\l\ .
\eel{Rindlerboost} This implies that its value for the out-state is the same as for the in-state. The ``algebra" is
therefore,
\be\F^i_\inn(\tl\s)&=&\F^i_\out(\tl\s)\ ; \\
\big[\F^i(\tl\s),\,\F^j(\tl\s')\big]&=&0\ . \eel{scalaralg}

This, however, is important. The scalar fields \(\F^i(\tl\s)\) give the horizon some conserved degrees of freedom; there is
a local conservation law!  This local conservation law, however, differs from the local conservation laws in quantized gauge
field theories; in a quantized gauge theory, the locally conserved quantities always vanish. Here, they can take any set of
values. The \(\F\) fields for an in-state simply take the same values as for its corresponding out-state.

The role of the scalar field self-interaction, \(V(\mathbf\F)\), is the following. As long as these fields commute with all
other operator fields on the horizon, we must view them as Casimir operators. The quantity \be \mathbf{Z}(x^1,\,\cdots\,,\
x^n)=e^{\mathbf{W}(x^1,\,\cdots\,,\ x^n)}=\bra\mathbf\F(x^1)\,\cdots\,,\,\mathbf\F(x^n)\ket\ , \eel{scalarexpvalue}
describes the correlations for the values of \(\mathbf\F\) as they occur in the ``average" representation on the horizon.
Here, \(\bra\mathbf\F(x^1)\,\cdots\,,\,\mathbf\F(x^n)\ket\) is the standard Green function, defined at the points \(x^i\) on
the horizon. Since these points are all spacelike separated, the quantity in question can be computed as if we were in
Euclidean space rather than Minkowski space-time. This Green function, of course, depends directly on the values of the
scalar self-interaction \(V(\mathbf\F)\).

We imagine that the set of all states is finite (anticipating the fact that the entropy is finite), and that we must be
describing an arbitrary, generic, element of this set. The role played by the scalar field Lagrangian is to determine the
\emph{weight} of the values for the scalar fields when averaging over all black hole states. This is important, when we will
be describing the BEH mechanism, see Section \ref{Higgs.sec}.

If the scalar field \(\mathbf\F\) transforms non-trivially under some symmetry transformation, be it a global one or a local
one, then this symmetry may appear to be explicitly broken at the horizon, since the field takes a fixed value. This,
however, can be seen in a different light: since all values for the field \(\mathbf\F\) occur in the complete set of all
black holes, it may be more accurate to state that the black holes are degenerate under this symmetry: the entire black hole
transforms into another one under the symmetry transformation. If the symmetry group, or even its covering group, is not
compact, then a problem arises: the black hole appears to be in an infinite representation. Since its entropy is finite,
such symmetries are not allowed. This is Bekenstein's well-known argument\cite{Bekenstein} that black holes cannot observe
the additive conservation laws associated to non-compact symmetry groups.

When studying a finite region of the horizon, it is the surrounding scalar field that appears to break the symmetry
explicitly. In this case, the scalar field acts exactly as if it were a \emph{spurion}. Spurions\cite{spurion} were
introduced in the '50s and '60s to describe explicit symmetry breaking such as the breaking of isospin and flavor-\(SU(3)\).

\newsec{The BEH mechanism}\label{Higgs.sec}

Let us now introduce a complex Higgs field \(\f(x)\), replacing the Lagrangian \eqn{MaxwL} by \be\LL(x)=-\quart
F_{\m\n}F_{\m\n}-(D_\m\f)^*D_\m\f-V(\f,\f^*)\ , \eel{HiggsL} where \(D_\m=\pa_\m+ieA_\m\), and \(V\) must be
gauge-invariant, typically \(V=\half\l_H(\f^*\f-F^2)^2\), so that \(\bra\f\ket\ra F\).

 In Ref.~\cite{GtH96}, we learned how to include the
effects due to the Higgs mechanism, henceforth referred to as the BEH mechanism\cite{BEH}. The primary effect of the Higgs
field is to add a mass term \(-\half M^2A_\m^2\) to the Lagrangian \eqn{MaxwL}, where \(M^2=2\l_HF^2\). This leads to a
similar mass term in \eqn{MaxwLJ}, so that the equation for the Green function \(f(\tl x-\tl x')\) in Eq.~\eqn{gaugejump} of
our algebra gets modified accordingly. This is because the Maxwell equations \eqn{Maxwequ} also receive this extra term. It
was subsequently noted that, apparently, local gauge invariance gets lost.

We can now understand in a more detailed way how this comes about. The scalar Higgs field, \(\f\), fluctuates around the
dominating values \(|\f|=F\). This breaks local gauge symmetry, but here, the breaking appears to be an explicit one, not a
spontaneous one, since the field \(\L(x)\) is a scalar field, not a vector field. Now, we connect this observation with our
treatment of the scalar field in Section \ref{scalar.sec}. The scalar field acts as a spurion. The algebra of
Eqs.~\eqn{rlcomm}, \eqn{rlinout}, and \eqn{rloutin} is kept unchanged, but Eq.~\eqn{gaugejump} is replaced by \be
(\tl\pa_\s^2-2e^2\f^*(\tl\s)\f(\tl\s))\,f(\tl\s-\s')=-\d^2(\tl\s-\s') \ . \eel{BEHjump} Thus, it is noted that the entire
scalar field, not just its vacuum expectation value, appears in the algebra.

\newsec{The transverse gauge field components} \label{transverse.sec}
Having learned what the role is of scalar fields in our algebra, we now return to the transverse vector field components,
\(\tl A(x)\). It is important to note that, just as the scalar fields, these components of the vector field also remain
invariant under a Rindler time boost. Clearly, they should be treated in the same way as our scalar fields. For instance,
the Higgs field must appear with the transverse covariant derivative in the Lagrangian \eqn{scalarL}, and indeed, the
transverse vector field plays a role when, in the previous Section, we rotate \(\f\) towards the positive real axis: \(\f\ra
F\) everywhere on the horizon. In a statistical sense, local gauge invariance is restored.

In summary: all field components that are invariant under the Rindler time translation, \be x^+&\ra& \l\, x^+\ , \nn
x^-&\ra& x^-/\l\ , \eel{Rindlertime} act as \emph{spurion fields} in the algebra; they are time translation invariant on the
horizon. The fields \(A_+(x)\) are associated with the \emph{in}-states, and \(A_-(x)\) with the \emph{out}-states, so that
they generate the algebra \eqn{rinlout}, \eqn{inoutcomm}, but with modifications such as \eqn{BEHjump}, due to the spurions.

\newsec{Conclusion. The algebra from the Lagrangian}
\def\SM{\mathrm{SM}}
We started with a \(U(1)\) Lagrangian with scalars: \be\LL^\SM(x)=-\quart F_{\m\n}F_{\m\n}-(D_\m\mathbf\F(x))^*
D_\m\mathbf\F(x) - V(\mathbf\F,\,\mathbf\F^*)\ , \eel{LAbel} with the usual definitions for an Abelian field strength
\(F_{\m\n}\) and the covariant derivatives \(D_\m\) for a set of scalar fields \(\mathbf\F\).

The complete set of all possible black hole states form a distribution in the space of \(\mathbf\F\)- and \(\tl A\) values.
The moments of this distribution must be described by the quantum vacuum correlation functions
\(\bra\mathbf\F(\tl\s^1)\cdots\,\tl A(\tl\s^n)\ket\). The relation between the fields \(A_\pm(\tl\s)\) and the sources
\(\r_\inn,\ \r_\out\) are then given by solving the classical Euler-Lagrange equations, using the given values for
\(\mathbf\F(\tl\s)\,\) and \(\tl A(\tl\s)\), for the action \be S=\int\dd^4x\LL^\SM(x)+\int\dd^2\tl\s\Big(\r_\inn(\tl\s)
\l_\inn(\tl\s)-\r_\out(\tl\s)\l_\out(\tl\s)\Big)\ , \eel{SAbel} where \be A_+(x)=\d(x^+)\l_\out(\tl\s)\qquad
\hbox{and}\qquad   A_-(x)=\d(x^-)\l_\inn(\tl\s)\ . \eel{Epots}

Since \be\pa_-A_+=0\ ,\qquad\pa_+A_-=0\ , \eel{dpmA} the longitudinal derivatives of \(A_\pm\) will disappear from the
kinetic part of \(\LL^\SM\). The action for \(\l_\inn\) and \(\l_\out\) will reduce to
 \pagebreak[1]
\be S&\ra&\int\dd^4x\Big(-\tl\pa A_+\tl\pa A_- -A_+(q\mathbf\F)^*A_-q\mathbf\F-A_-(q\mathbf\F)^*A_+q\mathbf\F \Big)\nn &&
+\int\dd^2\tl\s\Big(\r_\inn\l_\inn-\r_\out\l_\out\Big) \nn &=& \int\dd^2\tl\s\Big(-\tl\pa\l_\out(\tl\s)
\tl\pa\l_\inn(\tl\s)-2(q\mathbf\F)^* q\mathbf\F(\tl\s)\,\l_\out(\tl\s)\l_\inn(\tl\s)\nn &&
+\r_\inn(\tl\s)\l_\inn(\tl\s)-\r_\out(\tl\s)\l_\out(\tl\s)\Big)\ , \eel{Lrhol} where \(q\) is the charge matrix for the
scalar fields \(\mathbf\F\).  This gives the relations
\be\r_\inn(\tl\s)=-\tl\pa_\s^2\l_\out+\cdots,\qquad\r_\out=\tl\pa_\s^2\l_\inn+\cdots\,, \eel{rdl2} where the ellipses refer
to the contributions of the spurion fields \(\mathbf\F\) on the horizon. The algebra is now generated by Eq.~\eqn{rlcomm}.

The next step to be taken is to generalize this result to the case of non-Abelian vector fields and fermions. There are
several avenues to be explored, and we postpone this to a next publication.


\appendix


\newsec{Unitarity}\label{unitarity.app}
When expectation values of operators \(\OO(x)\) are computed in the presence of Hawking radiation, where \(x\) is in the
region of the Universe from which information can reach the outside world, one finds the outcome to have the general form
\be \bra\OO(x)\ket=\Tr(\r\,\OO(x))\ ,\qquad \r=e^{-\b H}\ ;\eel{expvalue} where \(H\) is the Rindler Hamiltonian and \(\b\)
the inverse Hawking temperature. At first sight, this result does not seem to depend on the state \(|\j\ket_\inn\) of
whatever it was that collapsed into a black hole, so it was initially suspected that pure quantum states \(|\j\ket\bra\j|\)
could, at least in principle, evolve into mixed quantum states \(\r\). Indeed, why could one not replace the familiar
scattering operator \(S\) by an operator \(\$\) that describes just any linear mapping of operators \(\r\) onto other such
operators\cite{Hawking87}? Instead of the Hamiltonian \(H\) of conventional Quantum Mechanics, which forces a density matrix
\(\r(t)\) to evolve according to \be \dd\r(t)/\dd t=-i[H,\,\r(t)]\ ,\eel{pureHam} one could postulate a more general linear
function of \(\r(t)\): \be \dd\r(t)/\dd t=-i\mathbb{H}\,\r(t)\ ,\eel{mixedHam} where \(\mathbb{H}\) acts as a general linear
operator.

\(\r\) is an element of a product Hilbert space, \(\r(t)\in\HH_I\otimes \HH_{II}\,\), so if we write \be \qquad
\r(t)=\r_{ij}|\j_i\ket_I\,\bra\j_j|_{II}\ ,\eel{mixedbasis} then \(\mathbb{H}\) can be written as \be
\mathbb{H}=H_I-H_{II}+H^\intt\ ,\eel{Hamexpand} where \(H\) is the ordinary quantum Hamiltonian and \(H^\intt\) describes
the extent by which \(\mathbb{H}\) deviates from the purely quantum mechanical one. The expectation value of an operator
\(\OO\) is \(\mathbb{H}\) \be  \mathbb{H} \bra\OO\ket=\bra\j|\OO|\j\ket=\Tr(\r\,\OO)=\bbs\bra \OO\bbs | \r\bbs\ket\ ,
\eel{expvalues} where the double bra-ket notation refers to inner products in the product Hilbert space \(\HH_I\otimes
\HH_{II}\), and \(\bbs|\OO\bbs\ket\) stands for \(\sum_n\OO_n|\j_n\ket\bra\j_n|\), where \(|\j_n\ket\) are the eigenstates
of \(\OO\) with eigenvalues \(\OO_n\). There are two important conditions on \(\mathbb{H}\):\bi{$i$} \emph{probability
conservation}: \(\mathbb{H}\bbs |\mathbb{I}\bbs\ket =H^\intt \bbs |\mathbb I\bbs\ket =0\) , and \itm{$ii$} \emph{energy
conservation}: There exists an operator \(E\), closely related to \(H\), and also bounded from below, such that \(\bbs\bra
E'\bbs|\mathbb{H}\bbs |E\bbs\ket=0\ \) if \(\  E'\ne E\ .\)\ei Here, \(\mathbb I\) is the identity operator. The first
condition seems to be not too difficult to meet, when we write down some trial operator for \(H^\intt\).

The second requirement, however, will not be easy to meet. If \(E=H\), it requires that \(H^\intt\) vanishes on all pure
energy eigenstates, a condition that seems to violate locality. An argument that seems to be more compelling physically is,
that \(H^\intt\) would tend to mix states from \(\HH_I\) and \(\HH_{II}\). If the first two terms of Eq.~\eqn{Hamexpand} are
viewed as an ``unperturbed" Hamiltonian, then we see that states in a universe where all energies are positive, interact
with states in a universe where all energies are negative. Energy conservation, as it follows from time translation
invariance, then tells us that energy exchanges are such that energy can be added to \(H_I\) while the same amount is also
\emph{added} to \(H_{II}\). The \emph{phase space} of all states where both \(H_I\) and \(H_{II}\) are larger, is much
bigger than phase space where both energies are smaller. This implies that, on average, energies in both universes will
rapidly increase, and this indicates a severe instability of the energy. Energy conservation is a problem.

A side remark here is that energy non-conservation or instability may not seem to be a problem in other cases of quantum
decoherence, described by density matrices. Recall, however, that in most models decoherence takes place as a consequence of
interaction with some external heat bath. In that case also, there will be energy exchange, leading to a thermal equilibrium
where the energy of the sub-system in consideration may be much more than at the starting point. It is important then to
realize that, in contrast, energy conservation should be essential for a system such as a black hole, to couple to a
gravitational field.

Imagine now that all these problems can nevertheless be solved, while still having an evolution law of the form
\eqn{mixedHam}. A more important and relevant question is then what the fundamental laws may be expected to be like. A
deterministic theory would allow for probability functions \(W_i(t)\), evolving according to a law \be\dd W_i(t)/\dd
t=\tl{\mathbb{H}}_{ij}W_j(t)\ ,\qquad \tl{\mathbb{H}} = -i\mathbb{H}\ , \eel{determ} while probability conservation would
demand \be{\dd\over\dd t}\sum_i W_i(t)=0\ \ra\ \tl{\mathbb{H}}\bbs|\mathbb I\bbs\ket=0\ ,\eel{probcons} equations reminding
us of the similar requirements for \(\r(t)\). However, a deterministic theory \emph{also} allows us to write equations of
the form \be{\dd \over\dd t}|\j(t)\ket=-i H|\j\ket\ , \eel{detSchro} where \(|\j\ket\) stands for an element of a Hilbert
space spanned by the \emph{equivalence classes} of ontological states. In the terminology of Ref\cite{GtH04}, we have the
somewhat counter intuitive situation that the evolution equation directly for the density matrix \(\r\) could follow for a
local, deterministic theory, whereas ``quantum purity" would be restored only if we replace the ontological states by
equivalence classes of states. In the author's view of the likely origin of the quantum mechanical nature of our world,
quantum mechanical purity is automatically restored, provided one employs the appropriate basis of states.

Hawking, in Ref.\cite{Hawking87}, mentions four possible remedies of the black hole information problem, of which only
option \# 3 resembles a bit the view taken here, except that he assumes the black hole information only to re-emerge just
before it collapses completely, a possibility that he rightly rejects. Curiously, the assumption taken here, that
information is re-emitted very shortly after it was absorbed, was not envisaged by him at that time.


\end{document}